# The Moving Beam Diffraction Geometry: the DIAD Application of a Diffraction Scanning-Probe

Authors

**Alberto Leonardi[a]\*, Andrew James, Christina Reinhard[bc], Michael Drakopoulos[d], Ben Williams[a], Hans Dehyle[e], Jacob Filik[a], Liam Perera[a]\* and Sharif Ahmed[a]\***

[a]Physical Sciences, Diamond Light Source (United Kingdom), Diamond House - Harwell Science & Innovation Campus, Didcot, Oxfordshire, OX11 0DE, United Kingdom

[b] The University of Manchester at Harwell, Harwell Science & Innovation Campus, Didcot, Oxfordshire, OX11 0DE, United Kingdom

[c]Faculty of Science and Engineering, The University of Manchester, Oxford Road, Manchester, M13 9PL, United Kingdom

[d]National Synchrotron Light Source II, Brookhaven National Laboratory, Bldg. 741 P.O. Box 5000, Upton, NY, 11973-5000, United States

[e]Biomaterials Science Center (BMC), University of Basel, Hegenheimermattweg 167C, Basel, Basel-Stadt, CH-4123, Switzerland

Correspondence email: alberto.leonardi@diamond.ac.uk; liam.perera@diamond.ac.uk; sharif.ahmed@diamond.ac.uk

**Synoptic**

We introduce the moving beam diffraction geometry as implemented at the Dual Imaging and Diffraction (DIAD) beamline at Diamond Light Source. We provide a quantitative assessment of the precision offered by this geometry and the nearest-neighbour calibration method.

**Abstract**

Understanding the interactions between microstructure, strain, phase, and material behavior is crucial in scientific fields such as energy storage, carbon sequestration, and biomedical engineering. However, quantifying these correlations is challenging, as it requires the use of multiple instruments and techniques, often separated by space and time. The Dual Imaging And Diffraction (DIAD) beamline at Diamond Light Source (DLS) is designed to address this challenge. DIAD allows its users to visualize internal structures (in 2D- and 3D), identify compositional/phase changes, and measure strain. It enables in-situ and in-operando experiments that require spatially correlated information. DIAD provides two independent beams combined at one sample position, allowing "quasi-simultaneous" X-ray Computed Tomography (XCT) and X-ray Powder Diffraction (XRPD). A unique functionality of the DIAD configuration is the ability to perform "image-guided diffraction", where the micron-sized diffraction beam is scanned over the complete area of the imaging field of view without moving the specimen. This moving beam diffraction geometry enables the study of fast-evolving and motion-susceptible processes and samples. Here, we discuss the novel moving beam diffraction geometry presenting the latest findings on the reliability of both geometry calibration and data reduction routines used. We provide a comprehensive quantitative assessment of the moving beam diffraction geometry implemented at the DIAD beamline, which will serve as a reference for beamline users. Our measures confirm diffraction is most sensitive to the moving geometry for the detector position downstream normal to the incident beam. The observed data confirm that the motion of the KB mirror coupled with a fixed aperture slit results in a rigid translation of the beam probe, without affecting the angle of the incident beam path to the sample. Our measures demonstrate a nearest-neighbour calibration can achieve the same accuracy as a self-calibrated geometry when the distance between calibrated and probed sample region is smaller or equal to the beam spot size. Indeed, we show the absolute error of the moving beam diffraction geometry at DIAD with typical calibration set-up remains below 0.01%, which is the accuracy we observe for the beamline with stable beam operation.

## 1. Introduction

Spatially correlated imaging and scattering information enables the study of complex systems across different spatial and temporal scales, such as energy storage materials, soil mechanics, bio-medical materials, and heritage or space mission samples (Besnard et al., 2023, 2024; Le Houx et al., 2023; Reed et al., 2024; Vashishtha et al., 2024). Extending from the opportunity of *ex-temporal* observations available at other instruments (Drakopoulos et al., 2015; King et al., 2016), the Dual Imaging and Diffraction (DIAD) beamline at Diamond Light Source provides inherent space and time correlation between structural and morphological information (Reinhard et al., 2021). Two independent synchrotron X-ray beams probe the sample with either imaging or diffraction data collection at a swap frequency rate of 20 Hz, making the corresponding exposure time the time-limiting factor. A rigid set of slits coupled with a Kirkpatrick–Baez (KB) mirror system is mounted on a pair of translation stages to achieve a high-resolution space correlation (Reinhard et al., 2021). The KB mirror focuses a 300 μm-by 300 μm region of the diffraction beam onto the sample with a spot size ranging from 50 μm by 50 μm to 15 μm by 5 μm. Translations of the KB system allow for selective area diffraction mapping by scanning of the focused X-ray micro-beam across the imaging volume. While the range of beam translation currently in use is 1.4 by 1.2mm, which matches the area captured by the PCO camera, the upgrade to an Andor Balor imaging detector is expected to extend the range to 2.0 by 2.0 mm. The live feed from the imaging stream, can be used to guide the position of the diffraction beam on the sample (Figure 1). Evolving processes can then be probed in space within the limitations of the exposure time for a diffractogram, typically ranging from 2 to 10 seconds (expected to significantly decrease with the ongoing Diamond ring upgrade), while the KB translation between opposite corners in the field-of-view is achieved in less than 0.5 seconds.

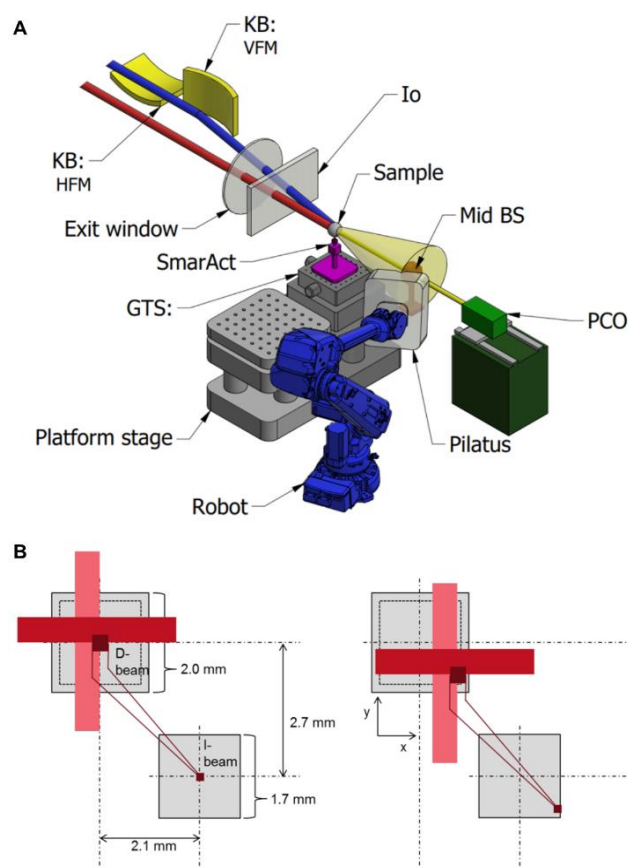

**Figure 1 Schematic of the moving beam scanning that is used on DIAD**. (**A**) Spatial configuration of the optical elements within the experimental hatch, showing the imaging (red) and diffraction (blue) beam paths. The Kirkpatrick–Baez (KB) focusing mirror refocuses the diffraction beam onto the sample position, which is spatially registered with the imaging field-of-view. (**B**) A set of fixed-aperture slits, mechanically coupled to the KB mirrors, selects a sub-region of the diffraction beam. The horizontal and vertical KB mirrors, equipped with fixed aperture plates, focus the dark red portion of the diffraction beam onto the sample. Due to the angular offset between the diffraction and imaging beam paths, the diffraction beam is focused across a field-of-view (~1.7 mm as per DIAD design), located approximately 2 mm from the projected position of the KB slits.

Scanning the diffraction beam across the sample breaks away from the conventional diffraction scanning approach adopted by other synchrotron beamlines, at which the diffraction beam stays stationary and the sample moves (Odstrcil et al., 2019; Yang et al., 2022). Movement of the X-ray beam in place of the sample avoids the need to coordinate high-speed precise movements of stages that support the sample and the surrounding sample environment with the need to avoid any sample perturbation, e.g., vibrations and composition concentration in liquids (Noell et al., 2020). Diffraction and radiography data can be collected without any movement of the sample enabling the study of processes highly sensitive to vibrations or kinetic effects, such as oxide-reduction reactions in a liquid solution. However, using a moving beam, the detector pixels measure a different location of the sample's reciprocal space at each beam position (see Appendix A). Movement of the diffraction beam introduces additional complexity into the standard diffraction data reduction pipelines. The area detector data reduction can further magnify instrument aberrations such as peak broadening, peak shifting, and peak splitting. Changes in the diffraction geometry affect the instrument's ability to quantify changes in the diffraction pattern accurately. Aberrations caused by either the reflection or transmission diffraction geometries are well known (Borchert, 2014; Cernik & Bushnell-Wye, 1991; Cullity, 1956; Guinier, 1956; Kriegner et al., 2015; Warren, 1990). Common instrument corrections for the analysis of scattering data collected with the Bragg-Brentano or Debye-Sherrer geometries are readily available in most line profile analysis software (Coelho, 2018; Perl et al., 2012; Toby & Von Dreele, 2013). The DIAD moving beam diffraction geometry is a new way of doing transmission diffraction. Therefore, an additional geometry calibration step is required for the data reduction process which directly affects the resolution of the moving beam diffraction geometry adopted by the DIAD instrument.

In this work, we present a systematic characterization of the aberrations associated with the moving beam diffraction geometry implemented at the DIAD beamline. The position and orientation of the diffraction detector relative to a reference X-ray source beam are calibrated by fitting the diffraction pattern from a standard NIST sample. Hence, the diffraction geometry for an arbitrary KB mirror motion position (KB-position) away from the reference is either extrapolated based on the KB stage displacement or directly assigned the geometry from the nearest-neighbour reference configuration. Here we discuss the extrapolation and nearest-neighbour geometry calibration reduction routines as they are implemented in the Data Analysis WorkbeNch (DAWN) software (Filik et al., 2017). Our results demonstrate uncertainties in the diffraction geometry make the extrapolation approach less accurate than the approximation introduced by the nearest-neighbour approach if a suitably dense grid of reference beam positions is calibrated. We finally present a systematic characterization of the diffraction resolution achieved by the DIAD instrument for different area detector arrangements and grid density of reference beam positions.

**2. Methods**

## 2.1. Diffraction geometry calibration

At synchrotron sources, diffraction geometries are adjusted to work with complex instrument constraints leading to tailored configurations for individual experiments. Compared to conventional laboratory diffractometers, the choice of a detector position, sample type, and sample positioning is unique to each experiment. The most accurate assignment of the observed scattering momentum is then achieved via instrument calibration with a known standard sample performed at the start of every experiment. This problem led to the accurate investigation of different calibration strategies of a diffraction geometry using flat panel detectors (Hart et al., 2013). Ideal background models are now well established (Caglioti et al., 1958; Cheary et al., 2004). The calibration of a single scattering image can broadly be broken down into the following steps: point-finding, outlier rejection, indexing, and diffraction geometry optimization. An exhaustive discussion of each of these steps is beyond our scope. Here we focus on the key areas required to understand the challenges introduced by the movement of the beam implemented at DIAD.

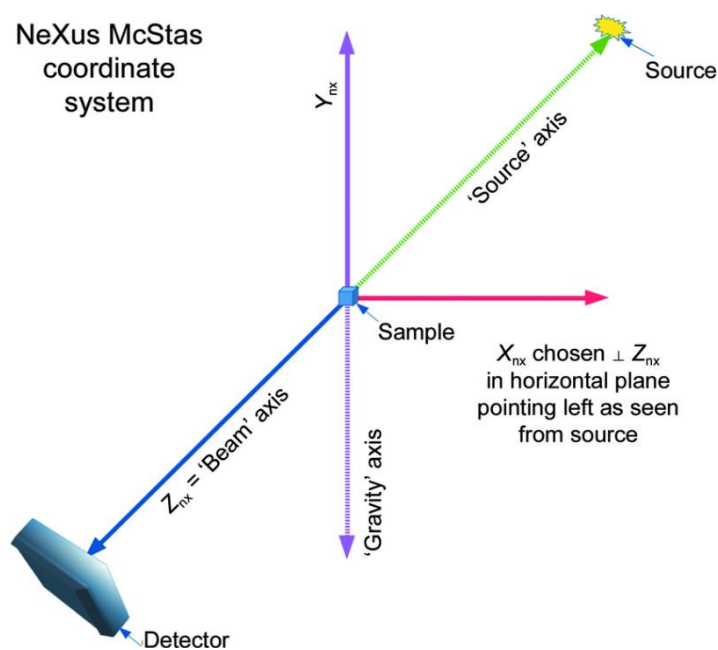

**Figure 2 Coordinate reference system at the DIAD beamline.** The McStas coordinate system convention is shown with y-axes opposite the gravity direction and z-axes aligned with the imaging beam. The area detector is shown in the real space configuration, and the observed portion of the Ewald sphere is mapped in the reciprocal space. Taken from Bernstein et al. (2020).

In the DIAD beamline, we fine-tune harmonic rejection and monochromator mirrors' orientations (i.e., Pitch and Roll) to align the imaging and diffraction incident beams with the axes of rotation of the tomography stage, which is otherwise kept stable over time in position and direction thanks to a set of high-precision active step motors. The intersection of the imaging beam with the rotation axes then defines the origin of the beamline reference system as co-located with the sample position. Bragg's law describes the kinematic diffraction, which relates the scattering angle 2θ with the structure symmetries of the periodic arrangement of atoms in a crystal. Constructive interference is observed for $n\lambda = 2d \sin \theta$ (Bragg & Bragg, 1913), where $d$ is the interplanar distance, $\lambda$ the wavelength, $n$ the reflection order, and the scattering angle $2\theta$ is defined as the angle between the incident beam and the center of scattering to the detector pixel direction. The sample is then the origin of the reference system for the definition of the diffraction geometry (He, 2009). Hereinafter we adopt the McStas standard to describe the reference system (Figure 2), although alternative options are also commonly used (Bernstein et al., 2020; Lefmann & Nielsen, 1999; P. Willendrup et al., 2004, 2014; P. K. Willendrup & Lefmann, 2020, 2021). The McStas system describes the incident beam vector as the cardinal direction $\mathbf{z} = [0,0,1]^T$, and assigns the $\mathbf{y} = [0,1,0]^T$ direction as opposite to gravity. The detector is then described by its normal direction $\hat{\mathbf{n}}$, pointing away from the sample, and the vector $\mathbf{d}$ at which the detector plane intercepts the direct beam.

It is worth noting that the moving beam geometry of DIAD yields inevitable definition discrepancies. To ensure that both imaging and diffraction center of scattering coincide at the sample position, the diffraction incident beam's real direction has a shallow angle of ~0.2° relative to the imaging incident beam, which is used to align the beamline optics and registers the beamline cartesian reference system. The real-space detector-pixel position, $\mathbf{p}$, and scattering momentum vectors, $\mathbf{q}$, are related through the Ewald sphere construction (Ewald, 1921). In the Born kinematic approximation of the elastic scattering,

$$\mathbf{q} = \mathbf{s} - \mathbf{s}_0, \tag{1}$$

where $\hat{\mathbf{s}}_0$ and $\hat{\mathbf{s}}$ are, respectively, the directions of the incident and scattered rays, and $|\mathbf{s}_0| = |\mathbf{s}| = 2\pi/\lambda$. Hence, in the case where the detector normal, $\hat{\mathbf{n}}$, is not orthogonal to the propagation direction of the scattered beam $\hat{\mathbf{s}}$,

$$|\mathbf{q}| = Q = \frac{4\pi \sin \theta}{\lambda}, \tag{2}$$

, and

$$\mathbf{p} = \alpha_s \hat{\mathbf{s}}, \tag{3}$$

with

$$\alpha_s = (\hat{\mathbf{n}} \cdot \mathbf{d})/(\hat{\mathbf{n}} \cdot \hat{\mathbf{s}}). \tag{4}$$

the distance from the scattering origin to the scattered rays intercept with the detector plane (Figure 2).

During the point-finding step, i.e., identification and location of Bragg diffractions, the area detector is mapped into the reciprocal space based on the observation of the Debye-Scherrer cones by the corresponding diffraction rings from a powder sample, or the reflection spots from a single-crystal sample. A National Institute of Standards and Technology (NIST) certified reference material is used to ensure the lattice parameters are known with high accuracy, with the assumption of a cubic structure.

A full calibration of the diffraction geometry requires the solution of seven degrees of freedom: the photon energy of the incident beam ($E = hc/\lambda$), two coordinates of a point in the detector plane mapped into the reciprocal space, the sample-to-detector distance, and three Euler angles that define the orientation of the detector plane. Different models for calibration vary by sample type (e.g., single crystal *versus* powder), parameterization of the detector geometry (beam center intercept *versus* PONI (Kieffer et al., 2020), and point- or ellipse-based refinement of the diffraction rings. It should be noted that ellipse-based fitting routines require powder reference standards and full rings to be visible on the detector to limit the uncertainties of the ellipse parameters (Hart et al., 2013). DIAD's dual-technique design limits access to small-angle scattering and the azimuthal range to at most slightly less than 90°. Indeed, the imaging apparatus, scintillator housing and rail optics, and the collision protection systems (Figure 1a) prevent the diffraction detector from being placed directly downstream of the beam path whilst capturing imaging data. This limits the azimuthal angular region available for calibration, as well as for measurement. Partial ring position refinement makes it difficult to decouple the detector distance from the energy and beam center, especially at large tilt angles. Hence, the photon energy of the instrument is measured independently of the final detector position.

**2.2. Challenges from the moving beam diffraction geometry**

Changes in the beam position alter the origin of the diffraction geometry reference system, affecting the scattering momentum vector observed by any detector pixel (Figure 3). The inaccurate calibration of the diffraction data alters the interpretation of directional dependent measurements, such as strain and texture, and the interpretation of peak broadening effects in the azimuthally integrated data due to the errors of the beam center. The movement of the diffraction beam invalidates the use of a single calibrated reciprocal space mapping of the pixels.

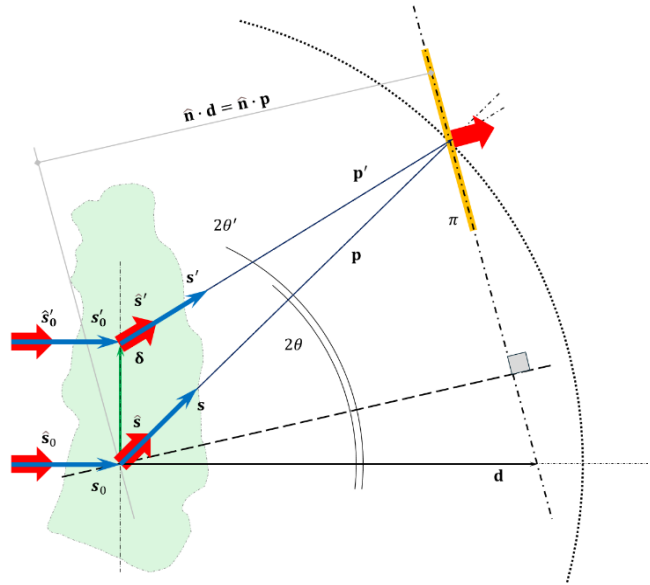

**Figure 3 The moving beam diffraction geometry.** The beam ($\hat{\mathbf{s}}_0$) incident to the sample (green shadow) is shifted by $\boldsymbol{\delta}$ to and $\hat{\mathbf{s}}_0'$ while the detector position and orientation is constant (yellow plane $\pi$). Although the detector distance $\mathbf{d}$ from the sample is not affected, the detector pixels will experience a relative change of their position from **p** to **p'** because of the alteration of the diffraction geometry. For a given detector pixel, the scattered vector, $\hat{\mathbf{s}}$ changes to and $\hat{\mathbf{s}}'$ affecting the angle relative to the incident beam (i.e., $2\theta$ to $2\theta'$). This causes a shift in the recorded $Q$ parameter.

A translation $\boldsymbol{\delta}$ of the reference system origin causes the translation of the scattering vector **s** associated with a given calibrated detector-pixel position **p**. Hence, after translation, the detector-pixel position in the translated reference system will be,

$$\mathbf{p}' = \mathbf{p} - \boldsymbol{\delta}. \tag{5}$$

The new scattering vector,

$$\mathbf{s}' = \frac{2\pi}{\lambda} \frac{\mathbf{p}-\boldsymbol{\delta}}{|\mathbf{p}-\boldsymbol{\delta}|}, \tag{6}$$

determines, then, the new scattering momentum $\mathbf{q}' = \mathbf{q} + \Delta\mathbf{q}$, with the detector-pixel calibration error (see Appendix B for the derivation),

$$\Delta\mathbf{q} = \frac{2\pi}{\lambda} \frac{(\alpha_s - |\mathbf{p}-\boldsymbol{\delta}|)\hat{\mathbf{s}} - \boldsymbol{\delta}}{|\mathbf{p}-\boldsymbol{\delta}|}. \tag{7}$$

A robotic arm allows for adjusting DIAD's detector configuration in space (Reinhard et al., 2022). In particular, the detector's outward normal $\hat{\mathbf{n}}$ and the sample-to-detector distance $r$ can be specified in a spherical reference system as,

$$\begin{cases} \hat{\mathbf{n}} = \begin{bmatrix} \sin\alpha \\ \sin\beta\cos\alpha \\ \cos\beta\cos\alpha \end{bmatrix} \\ r = \mathbf{d}\cdot\hat{\mathbf{n}} \end{cases} \quad (8)$$

with $\alpha$ and $\beta$ the cartesian angles with the McStas reference system.

Figure 4 compares the effect of a 0.5 mm diffraction scanning beam offset in the x-direction, i.e., horizontal normal-to-the-beam stream, which is about half of the imaging field-of-view, for a point detector on the forward scattering sphere octant with the sample-to-detector distance of 330 mm and 600 mm. The change of azimuthal angle causes a scattering momentum error proportional to the component of the beam offset, $\boldsymbol{\delta}$, orthogonal to the pixel detector position, $\mathbf{p}$, relative to the sample-to-detector distance, $\alpha_s$. Although the error diverges in the small-angle region, in the wide-angle region away from the direct beam the relative scattering momentum error is ~0.1% (i.e., $1\times 10^{-3}$) or ~0.05% (i.e., $5\times 10^{-4}$) for the sample-to-detector distance of 330 mm and 600 mm, respectively.

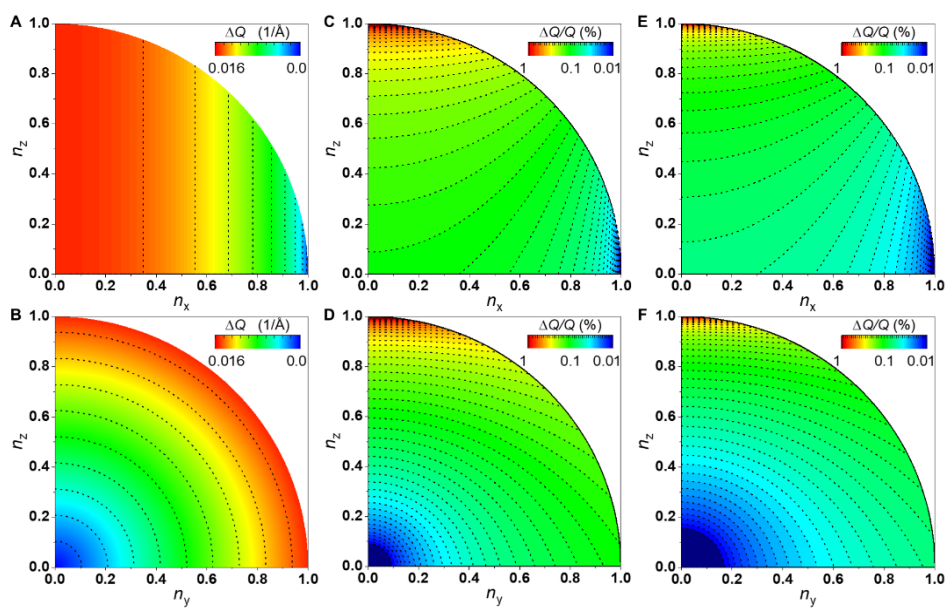

**Figure 4 Effect of moving the diffraction beam on the scattering observation.** Scattering momentum error $\Delta Q = |\Delta \mathbf{q}|$ for a detector pixel on the first sphere octant arising from a 0.5 mm beam offset along x. The error is shown for the sample-to-detector distances of 330 mm (**A-D**) and 600 mm (**E-F**), assuming a 20 keV radiation (0.619924 Å) and the McStas reference system (i.e., the incident radiation is parallel to z). The relative error $\Delta Q/Q$ (**C-F**) is independent of the radiation energy. The data in the upper and bottom row differ only for the projection base $n_x$-$n_z$ (**A**, **C**, and **E**) *vs.* $n_y$-$n_z$ (**B**, **D**, and **F**). Note that the scattering momentum $Q$ varies from 0 Å$^{-1}$ to ~14 Å$^{-1}$ with $n_z$ from 1.0 to 0.0 (i.e., $Q$ is independent of $n_x$ and $n_y$).

Additionally, given in the optimal configuration the pixel sensors in a flat-panel detector are arranged on a surface tangent to the sphere centered at the sample, the projection error introduced at each pixel strongly depends on the combined effect of the beam offset and each pixel's location on the detector.

This effect magnifies with the detector orientation such that none of the sensors is tangential to the sphere. It is, therefore, imperative to assign the correct detector orientation to avoid any diffraction geometry aberration and measure accurately the magnitude and direction changes of the probe.

Compared to traditional strain gauges that can achieve a resolution of about $1 \times 10^{-5}$, diffraction methods are expected to offer a strain resolution of one order of magnitude higher, *circa* $5 \times 10^{-6}$, to suitably access information on the structure deformation. Hence, the beam offset causes an error significantly larger than the resolution required for mechanical strain analysis in materials. A new calibration protocol is required to recover the expected accuracy, making the moving beam an independent diffraction geometry compared to other established alternatives.

**2.3. Calibration of the moving beam diffraction geometry**

The accuracy of any diffraction information computed from area detector observations, i.e., structure and microstructure properties, relies on the ability to resolve scattering angles and energy of the incident radiation with high reliability. Methods for high-resolution calibration of area detectors leverage multiple images or complex adjustable detector positioning to improve estimates of scattering geometry parameters (Hart et al., 2013; Horn et al., 2019; Wright et al., 2022). Here we use an array of diffractograms to overcome the ambiguities produced by the rotation of the conic section about the beam axis and the full detector orientation for a single reference image. We calibrate a first single image to be used as seed for further automatic calibrations. The data collected for a grid of diffraction beam offsets with known rigid translations of the focusing is then independently calibrated. The true space configuration of the detector could be reconstructed based on the changes observed in the detector origin estimated from the changes in pixel position of the diffraction rings. The effect of the offset of the incident beam should then be determined by simple extrapolation.

In contrast to stationary experiments, the correct seed detector configuration is key to the accuracy and confidence of the extrapolation procedure. It is worth noting that the extrapolation process assumes that the beam orientation of the incident beam is independent of the beam position. At DIAD the rigid translation of the focusing Kirk-Baez (KB) mirror with two high-precision step motors realizes the offset of the diffraction probe. These motions are built orthogonal to the beamline general design z-axes, which is generally not aligned with the diffraction nor with the imaging beams. This causes the moving beam operation to work with unavoidable unknown geometric alterations on the incident beam in addition to the idealized pure translation. Because of these constraints and the overall approximations involved with the extrapolation process, our default approach is to rely only on the directly calibrated array of data sets.

Nearest neighbour routines assign the instrument geometry calibrated of the grid's diffractograms collected for a standard sample with the closest KB-position to the experiment dataset. We assume then the corrected calibration of the beam movement remains valid regardless of the sample, as long as there is no change of diffraction detector and beamline optics configuration (i.e., orientations and positions). Moreover, centering of the sample on the tomography stage rotation axis ensures that sample movements away from the calibrated position can be reduced to less than the resolution of the imaging branch 1.6 µm, i.e., three times the imaging pixel size of 0.54 µm.

### 2.4. Application of the moving beam geometry in experiments

Compared to traditional Bragg–Brentano and Debye–Scherrer geometries, which are respectively optimized for flat-plate and capillary phase-homogeneous samples, the moving beam geometry is optimized for the study of spatially phase and morphology inhomogeneous samples. The data collection in transmission-mode enables the use of non-standard sample geometries, which is essential for the core purpose of the moving beam: to spatially resolve structural variations across macroscopically inhomogeneous systems. The moving beam geometry requires samples to have a microstructure such that the crystal size yields a powder-like scattering from a probe volume of a few µm. Hence, the use at DIAD of a Kirkpatrick–Baez (KB) mirror system to dynamically adapt the beam focus according to the crystal size and the required spatial resolution of different samples. DIAD offers three standard set-ups with focused beam to an area of 5 µm by 15 µm, 25 µm by 25µm, or 50 µm by 50 µm, which are suitable for a wide range of fields of study, including mechanical engineering, geosciences, energy materials, and medical and biological systems. The trade-off for this limitation is the key advantage of the moving beam geometry compared to more traditional geometries: the opportunity to probe different well-defined regions of the samples without their perturbation by sample movement. This opens, particularly, to the study of processes affected by time-dependent alterations of the chemical components' concentration in a liquid environment, where sample motion would cause re-mixing and so alter the time evolution process. Examples are the study of crystallization phenomena from liquid or melted liquid bath, metal oxidation, degradation of ceramic and biological materials exposed to aggressive liquid environments (e.g., salty water and body fluid simulants), and chemical reactors.

With the region to be probed not being stable, the geometry calibration with a standard capillary geometry becomes impractical and a flat-plate sample in transmission mode aligned orthogonally to the imaging beam path is left as the best average approximation of a generic macroscopic shape sample. The calibration on a plane orthogonal to the imaging beam path provides the best balance between induced aberrations (e.g., sample shift, absorption, sample geometry, etc.) and access to local information. In particular, the effective center of scattering is shifted compared to the calibrated position either because of the sample shape or absorption inhomogeneity. As the actual center of scattering can change across the sample itself or sample by sample, the calibration for the 2D plane standard provides a reference point for sample dependent corrections to be applied at the stage of line profile analysis. Our typical calibration standards, with nominal thicknesses of 0.5 mm, offer reasonably accurate flatness despite challenges in powder packing. While static samples inherently suffer from reduced statistical averaging and increased preferential orientation effects compared to spinning geometries, these limitations are well-characterized and do not significantly impact geometry calibration. However, they do pose challenges for quantitative diffraction analysis, particularly in texture-sensitive applications. Compared to traditional diffractometer instruments, the typical sample thickness remains limited to less than 1 mm to fit in the imaging field-of-view and most of the sample geometry and composition aberrations can be corrected based on the imaging radiography or tomography information. Indeed, imaging provides not only the information for a possible full absorption correction, but also a measure of the sample shift from the calibrated center of scattering with a resolution of 0.5 μm.

## 2.5. Diffraction calibration configurations

Experimental constraints dictate the choice of detector position, standard reference material, and beam energy. Here we characterize the calibration accuracy and precision of the most common combinations to assess the reliability of the moving beam diffraction geometry of normal experiment visit operation.

### 2.5.1. Detector positions

The spherical system previously presented by Reinhard et al. (Reinhard et al., 2022) specifies the position and orientation of the detector. In this system, the center of the detector plane is imposed to be orthogonal to the direction of scattering making simpler the diffraction geometry calibration. We assess the quality of the calibration for two characteristic detector positions, one conventional downstream normal-to- and one inboard tilted-to-the-beam stream (Figure 5).

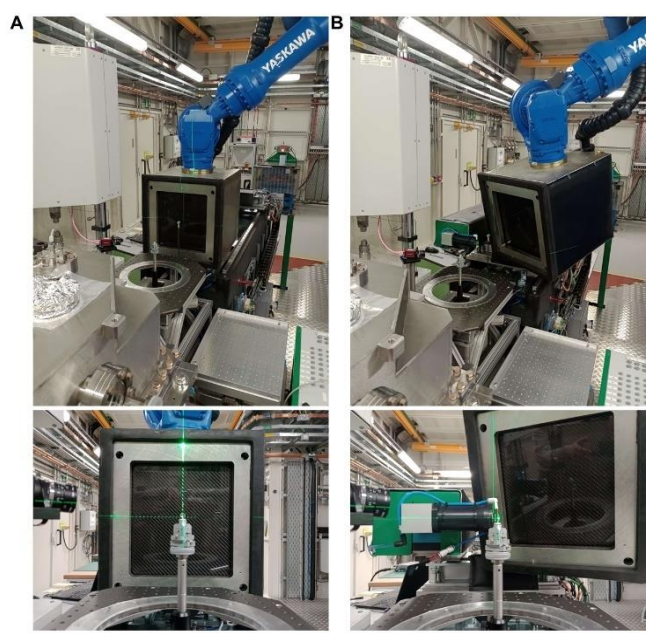

**Figure 5** Picture from the DIAD experimental hatch showing the detector positions (**A**) downstream normal-to- and (**B**) inboard tilted-to-the-beam stream.

With the detector position downstream normal-to- the beam stream the sample-to-detector distance is of ca. 330 mm and the pixel at the center of the 2D sensor panel is orthogonal to the scattering direction. Otherwise, the detector position inboard tilted-to-the-beam stream detector position is chosen with a sample-to-detector distance of ca. 300 mm and central detector pixel at α=27°, β=15°, and γ=0°. This position maximizes the ring section covered while still permitting imaging data to be acquired, representing a realistic experimental condition.

**2.5.2. Standard samples**

The standard reference materials 660b $LaB_6$ and 674b $CeO_2$ from the National Institute of Standards and Technology (NIST) are chosen to ensure a sufficient number of rings cover a large proportion of the detector plane and provide a suitable number of crystals to form complete powder rings without introducing excessive peak broadening. The cryo-cooled Si (111) double-crystal monochromator of the DIAD diffraction branch (DCM-1) operates in the energy range from 7 keV to 38 keV. At 25 keV and with the detector position in the downstream normal-to-the-beam stream detector position the first six and five innermost reflections of SRM 660b $LaB_6$ and SRM 674b $CeO_2$, respectively, are fully accessible. At energies higher than 25 keV, the rings of the SRM 660b $LaB_6$ come too close to ensure easy imaging processing, so the SRM 674b $CeO_2$ is preferable.

Calibration samples are prepared in 0.3 mm diameter borosilicate capillary and 0.5 mm thick ~2 cm wide polyimide flat-plate pocket samples. Although the capillary form allows the spinning to improve sampling statistics, the tubular shape limits the beam offset to a single scanning direction. Otherwise, the flat-plate shape enables the beam to offset along two independent scanning directions; but without the sample spinning the number of crystals contributing to the scattering signal is limited to those lying within the beam footprint through the sample. Samples are aligned with the axes of rotation using the imaging camera and the tomography stage. Although the samples are manually aligned with the gravity axes, the plane of the flat-plat holder is registered orthogonal to the imaging beam exploiting the general tomography stage (GTS) mounting the orthogonal x-z displacement motion on top of the rotation motor. Flat-plate samples are tilted until they are parallel with the imaging beam and then turned by 90° to be about orthogonal to the incident beams (Figure 6).

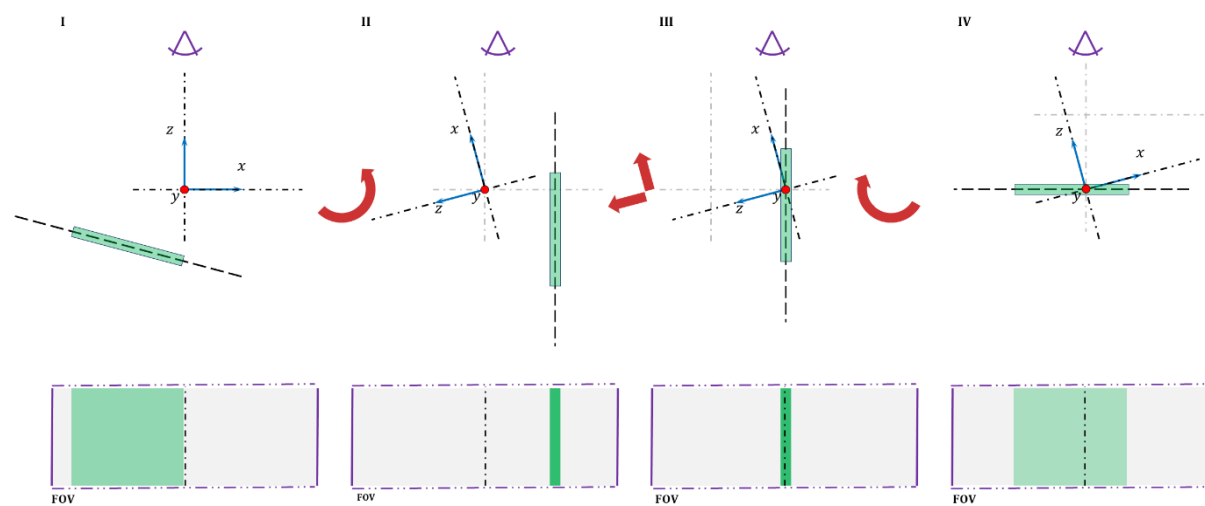

**Figure 6 Sample alignment with the tomography axes of rotation.** Samples are aligned with the general tomography stage axes of rotation exploiting the x-z motion being mounted above the rotation motor. After the flat-plate sample is oriented orthogonal to the FOV plane (I → II), it is shifted to the center of the FOV (II → III), which is aligned with the axes of rotation at the start of the beamline setup. Finally, a 90° rotation is applied to bring the plane of the sample orthogonal to the beam stream (III → IV).

### 2.5.3. X-ray energy

The beam energy is tailored to the experiment requirement based on pre-recorded Bragg angle and crystal-to-crystal distance tables. Despite this, the beam energy is estimated at the start of each new experiment session because it plays a crucial role in the diffraction geometry calibration as well as in the analysis of any diffraction data

Given at the start of the beamline setup the beam position at the sample is usually not yet synchronized with the KB motion, we use the diffraction signal from a standard flat-plate sample and the downstream normal-to-the-beam stream detector position to measure the energy of the incident beam. This position provides full ring coverage on the detector, and the direct beam lies close to the center of the detector. Several full reflections can be detected with a cone half opening angles of $2\theta \leq 19.5°$. The downstream configuration minimizes also the errors caused by uncertainty on beam position and the ambiguity between energy and sample-to-detector distance, both affecting the observed radius of the diffraction rings by the same factors. Samples are mounted and aligned with the rotation axis of the tomography stage before positioning the diffraction detector to exploit the imaging beam and detector camera. The beam energy value is optimized to match the pattern of the full diffraction rings based on the known spacing dictated by the standard material and an empirically optimized seed for the diffraction geometry and beam energy parameters. Indeed, it is well known that the results of automatic optimization routines are affected by the seed guess (i.e., the closer the seed is to the true solution, the more likely the optimization routine will converge to the latter).

Although this process allows for easy and time-efficient calibration of any energy ahead of normal operation, high-accuracy beam energy is otherwise achieved by detecting the absorption edge of a standard foil material. The Bragg angles of the DCM crystals are scanned and the integrated fluorescence intensity from the foil is measured across the imaging camera. The DCM is then set to the angle corresponding to the absorption edge of the foil (angle of maximum fluorescence). At DIAD, we use Zn, Mo, and Sn foils to independently identify the DCM angles for 9.66 keV, 20.00 keV and 29.20 keV energies.

**3. Results & discussion**

**3.1. Single image calibration accuracy**

The accuracy of the moving beam geometry is particularly sensitive to the accuracy of the reference calibration. Hence, uncertainties arising from the identified ring locations, limited ring coverage, and large diffraction spots can all degrade the quality of the correction.

The diffraction beam energy is calibrated at 25 keV and 30 keV with SRM 660b $LaB_6$ and SRM 674b $CeO_2$ samples, respectively. Between detector position movements the diffraction branch optics are not altered to ensure that the energy measurement made at the near-orthogonal downstream normal-to-the-beam stream detector position remains valid for the inboard tilted-to-the-beam detector position. Data are collected from both sample types using both detector positions (Figure 7). Diffraction data are collected using the Pilatus2M CdTe detector with a pixel size of 0.172 mm. A vertical scan of 120 points is taken on both samples using the KB-y movement only to ensure consistency between sample alignment, i.e., without sample movement. Acquisitions with the capillary samples are conducted with the stage rotating to maximize data quality, i.e., powder statistics. On flat-plate samples, an additional map of 121 grid points is measured per detector position using the KB movement orthogonal to the beam path. This consists of a 121×120 point grid centered with the field-of-view (FOV) of the imaging branch and covering a range of 1.2mm × 1.2mm with step sizes of 120 µm per each KB-motion.

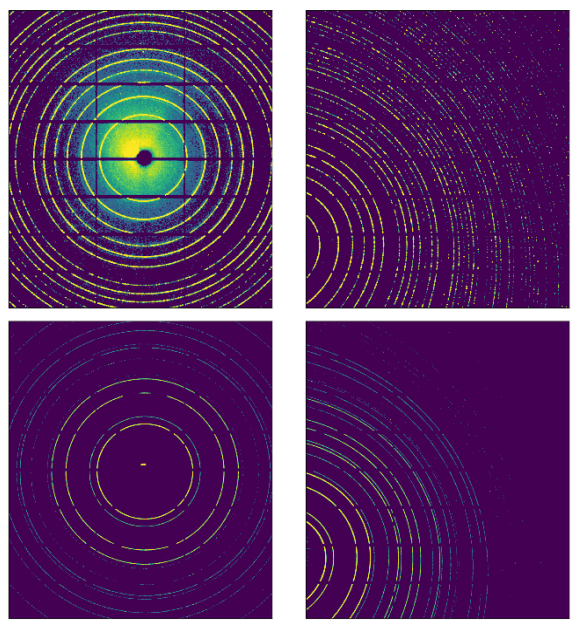

**Figure 7** Diffractograms from SRM 660b $LaB_6$ (upper row) and SRM 676b $CeO_2$ (bottom row), for the detector positions downstream normal-to- (left column) and inboard tilted-to- (right column) -the-beam stream.

A single crystal dataset is also studied to improve the accuracy of the detector roll angle, which plays a significant role in the correction procedure. The GTS rotation axis that holds the samples is specified as the $[0,1,0]^T$, assigning the tomography stage rotation axis anti-parallel to gravity consistent with the McStas coordinate system. A sphere of SRM 1990 $Al_2O_3$ is mounted on a 200μm MiTeGen loop and aligned to the rotation axis of a 1005 Huber goniometer. During the acquisition, the crystal is rotated about the reference axis at an angular rate of 0.1 deg/s with a detector exposure time of 1s. This ensures that individual Bragg reflections appear for multiple rotation steps, improving the reflection centroid refinement. In total, 16 datasets are investigated with the crystal orientated at multiple poses by manually rotating the sphere about the goniometer's axis. Assessment of the rotation center in the tomographic reconstruction confirms the rotation axis of the GTS is vertical in the imaging space and can therefore be used as a reliable reference for processing the single crystal datasets over the 16 poses.

### 3.1.1. Data analysis: calibration, reduction, and fitting

Individual frame calibration is performed using 500 detector-pixel points per diffraction ring. The radial region window to identify these points is computed based on the list of reflections of Table 1 and a rough seed manual calibration. The reference orientation and detector geometry are manually adjusted to provide a visual agreement between estimated and observed ring projections in the 2D detector image. Although the seed provides only a first approximation qualitative match, the calibration parameters are expected to resemble, unless for reference system convention change, the nominal configuration reported by the robotic arm that holds in position the Pilatus detector (Reinhard et al., 2022).

The detector orientation of the initial seed is converted to the equivalent 0°-roll configuration to obtain consistent calibration estimates of the diffraction beam offset frames. The scattering from a randomly orientated powder in a two-dimensional flat panel X-ray detector results in the same Debye-Scherer cones for different equivalent detector orientations. All these equivalent orientations differ by a simple change of the detector roll angle around the beam vector with the rotation centered at the beam center on the detector plane. Indeed, given the scattering cones share the same rotation symmetry axes, only two degrees of freedom among Yaw ($\alpha$), Pitch ($\beta$), and Roll ($\gamma$) are independent. Any detector orientation is then fully described by the two angles $\tilde{\alpha}$ and $\tilde{\beta}$. During individual frame calibration, the beam energy is fixed at the value determined from the calibration of the downstream normal-to-the-beam stream detector position, ensuring that the observation accurately reflects the uncertainties in the detector configuration only (see section 2.5.3 for more details).

**Table 1** Reflection sets for final position calibration and calibrated photon energy of the incident beam.

| SRM | Incident beam wavelength (Å) | Reflections | Representative peak FWHM (Å$^{-1}$) |
|---|---|---|---|
| 660b LaB$_6$ | 0.5008(16) | (1,0,0), (1,1,0), (1,1,1), (2,0,0), (2,1,0), (2,1,1), (2,2,0), (3,0,0), (3,1,0), (3,1,1), (3,2,0), (3,2,1), (4,1,0), (3,3,0), (4,2,0), (4,2,1) | 0.007 |
| 674b CeO$_2$ | 0.41456(11) | (1,0,0), (1,1,0), (1,1,1), (2,0,0), (2,1,1), (2,2,0), (3,1,0), (3,1,1), (2,2,2), (3,2,0), (4,0,0) | 0.009 |

The diffraction geometry is calibrated per each diffraction data set. Traditional Bragg intensity profiles are then computed *via* azimuthal integration using the Data Analysis Work beNch (DAWN) software (Filik et al., 2017). At Diamond Light Source, the DAWN software offers the ability to perform on-the-fly data reduction leveraging parallelization using high-performance computing (HPC) cluster. With DAWN, diffractograms are reduced to line plots using pixel-splitting integration. Other integration settings are adapted to the detector position, and sample material as reported in Table 2.

**Table 2** Azimuthal integration parameters based on detector position are used to provide a set of one-dimensional diffraction patterns as a function of detector angle.

| Detector position as $r$ (mm), | SRM | Number of | Radial Range start | Number of Radial |

| Detector position as r (mm), and α, β, γ (deg) | SRM | Azimuthal bins | and stop (Å$^{-1}$) | bins |
|---|---|---|---|---|
| 330, 0,0,0 | 660b LaB$_6$ | 360 | 1.094, 5.950 | 4200 |
| 300, 27,15,0 | 660b LaB$_6$ | 120 | 1.094, 5.950 | 4200 |
| 330, 0,0,0 | 674b CeO$_2$ | 360 | 1.80, 5.40 | 2700 |
| 300, 27,15,0 | 674b CeO$_2$ | 120 | 1.80, 5.40 | 2700 |

Peak fitting is applied to the data using a compound model of a Voigt profile in combination with a flat background. Individual peaks are isolated based on the expected peak positions and a suitable search range of 7σ (i.e., ±3.5 FWHM) to ensure additional correlations between the peak center uncertainty and window size are negligible (Withers et al., 2001). Peak fits are performed on several reflections with high multiplicity taken across various positions on the detector face. Peak fit reflections for the two materials are provided in Table 3.

**Table 3** Reflections used for assessment of calibration accuracy. Multiplicities have been added where multiple reflections overlap.

| Detector position as r (mm), and α, β, γ (deg) | SRM | Reflection family {hkl} | Multiplicity | Interplanar distance (Å$^{-1}$) |
|---|---|---|---|---|
| 330, 0,0,0 | 660b LaB$_6$ | 210 | 24 | 1.8593 |
| | | 211 | 24 | 1.6973 |
| 300, 27,15,0 | 660b LaB$_6$ | 210 | - | - |
| | | 211 | - | - |
| | | 310 | 24 | 1.3147 |
| 330, 0,0,0 | 674b CeO$_2$ | 311 | 24 | 1.6310 |
| | | 333 / 511 | 32 | 1.0411 |
| 300, 27,15,0 | 674b CeO$_2$ | 311 | - | - |
| | | 333 / 511 | - | - |
| | | 642 | 48 | 0.7229 |
| | | 733 / 555 | 48 | 0.7043 |

Single crystal geometry calibration is performed using Diffraction Integration for Advanced Light Sources (DIALS) software (Waterman et al., 2016). In the refinement, we use the certified ruby sphere crystal structure along with an initial estimate of the detector configuration. Prior to analysis, the DIAD standard NeXus file structure is reconfigured to conform to the NxMx standard format. The tilts of the two additional goniometer axes are determined from the analysis of the imaging datasets.

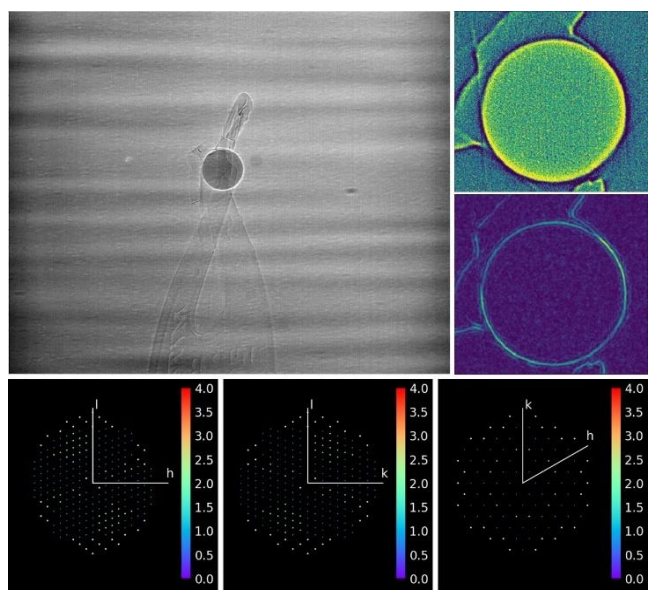

**Figure 8 Single crystal observation.** (upper-left corner) Radiography snapshot at 0 ° and (upper-row right) tomography reconstruction slice (top) and the same filtered (bottom) to measure sphere centroid. (bottom row) reciprocal space reconstruction with DIALS along three different zone axes.

A ruby sphere is aligned with the axes of rotation of the tomography stage and single crystal diffractograms are recorded at constant step tilt rotations (Figure 8). The small dimension of the crystal, below 0.2 mm in diameter, and the spherical geometry ensure minimum aberrations affect the diffraction data. The constant beam cross-section at different tilt angles removes the need for absorption correction. Results from DIALS analysis are in good agreement with expected values (see Table 4), such as the sample-to-detector distance of 365.447 mm. The detector Roll, Pitch, and Yaw angles, 24.87°, -20.46°, and -0.80°, respectively, are in good agreement with the corresponding values of 24.88°, -20.43° and -0.86° estimated by DAWN from analysis of powder diffraction data. However, the associated s.d.s of 0.015°, 0.034°, and 0.044° are significantly larger than the misorientation s.d. measured *via* DAWN refinements with known incident radiation energy, i.e. ~0.007° (See section 3.1.2 and Figure 10C for more details).

**Table 4** **Single crystal diffraction**. Structure solution with DIALS software.

| Nominal a, b, and c | 4.76080(29) Å | 4.76080(29) Å | 12.99568(87) Å |
| --- | --- | --- | --- |
| Estimated a, b, and c | 4.76301(3) Å | 4.76301(3) Å | 13.00170(13) Å |
| α, β, and γ | 90° | 90° | 120° |
| Fast axes | 0.909604 | -0.0196721 | 0.415011 |
| Slow axes | 0.125585 | -0.939136 | -0.319769 |
| Origin | 7.05702 | 265.668 | -318.778 |

Hence the single crystal diffraction is less reliable than the powder diffraction process to measure the accurate detector space configuration relative to the incident beam. This is reasonable as the single crystal diffraction method is based on a much smaller number of reflection data points. Moreover, single-crystal diffraction is known to be affected by uncertainty over detector position, centroids of the single-crystal reflections, sample centering and consequent procession effects, and stability of the beam position. Otherwise, the use of a large flat-area powder sample and the recognition of the diffraction rings from a 2D area detector provides a significant increase in statistics and accuracy of observations.

### 3.1.2. Moving beam calibration accuracy

From performing independent geometry calibrations for a two-dimensional grid of diffraction centers across the imaging field of view it is possible to investigate the assumption that the vector of the incident beam is stable across the full measurement area (i.e., $\hat{s}_0 \parallel \hat{s}_0'$). Figure 9 shows the relative peak position error estimated for the first four $LaB_6$ reflections as a function of the sample (Figure 9A and B) and diffraction beam (Figure 9C) displacements. The plots reveal a recurring correlation pattern between the observed peak positions and the azimuthal-sector angle of observation. This pattern is likely due to alignment errors in the detector modules within the Pilatus2M panel. However, these errors fall well within the precision uncertainty associated with peak positions for each azimuthal-sector angle. Notably, the deviation remains within a 0.2% margin, regardless of the motion considered or any theoretical corrections applied to the diffraction geometry for the moving beam. Notably, the same increase of error is observed either by moving the diffraction beam or the flat-plate standard sample in the x-direction, i.e., horizontal normal-to-the-beam. This demonstrates the error is mostly caused by imperfections of the sample shape, i.e., nonconstant thickness, and possible tilt or bending of the flat-plate support especially in the gravity y-direction. Another significant source for the observed error is the tilt between the incident diffraction and imaging beam resulting in a slight shift of the center of scattering along the beam path while screening the flat plate sample in the horizontal direction. Negligible improvement of results is achieved by compensation of the change of diffraction geometry accounting for the extrapolation of the beam shift and relative detector tilt as a function of the displacement of the center of scattering (Figure 9C second *vs.* third row). Indeed, the error more than doubles compared to self-calibrated datasets (Figure 9C first row). This demonstrates that the uncertainty of change of the diffraction geometry caused by the moving beam exceeds the accuracy of a single KB-position calibration.

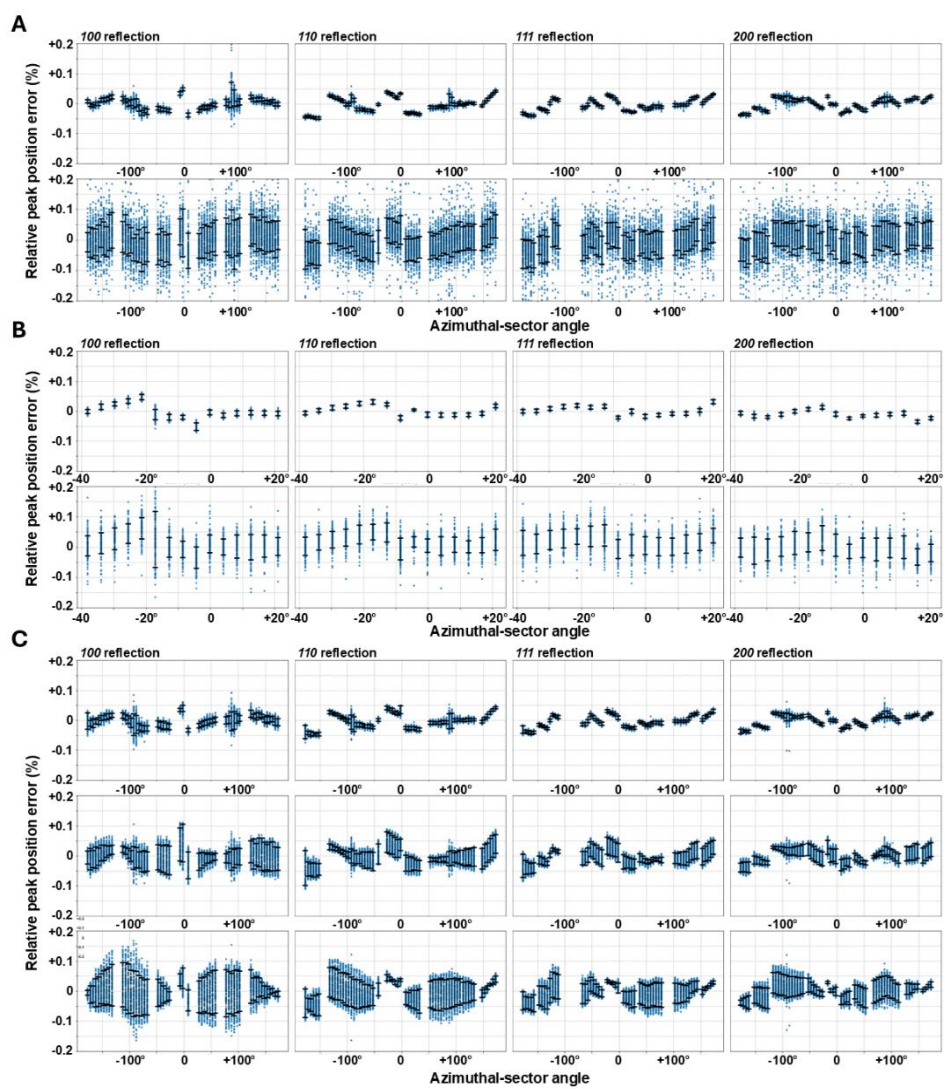

**Figure 9 Calibration accuracy.** Relative peak position error for the 100, 110, 111, and 200 reflections with the detector (**A**) downstream normal-to and (**B**, and **C**) inboard tilted-to-the-beam stream. Diffractograms were integrated over 5-degree sector of the azimuthal angle to assess precision and accuracy as a function of the projection angle across the detector panel. Diffraction measurements are repeated by applying a displacement in the gravity direction, i.e., tomography stage y axes, to the capillary sample (first row of A and B) and to a flat plate sample in the normal to the beam direction (second rows of A and B, and all C), i.e., tomography stage combined x and z axes. (**C**) the measurements are repeated for the same relative displacements of the center of scattering by moving the diffraction beam while the sample stays stationary. For this, the relative error is computed either using self-calibrated geometry (first row of C) or extrapolating the geometry correction for the sole beam shift (second row of C) or the combined beam shift and relative detector tilt (third row of C). Note that a relative peak position error of 0.05% corresponds to a relative interplanar distance error of ~0.0065% for a 1.8953 Å $d_0$ observed with a 25 keV X-ray.

Uncertainty of the detector orientation from a single diffractogram can be determined by taking the geodesic distance in $\mathbf{SO_3}$ (Haslwanter, 1995; Novelia & O'Reilly, 2015), i.e., the group of all possible rotations about the origin of the three-dimensional Euclidean space, between the individual detector orientation calibration and the corresponding orientation obtained from extrapolation through the orientation matrix construction. In DAWN two angles are used to define the orientation of the detector plane. The uncertainty in the final orientation is then determined representing the rotation matrices in the angle-axis formulation. The misorientation, $\varsigma$, between two orientation matrices, $\mathbf{R}_i$ and $\mathbf{R}_j$, is computed directly from knowledge of the two orientations. However, because of the scattering cone symmetry, it is possible that two orientations display large differences while being diffraction-equivalent data. Hence, adjustment of one of the rotation matrices to account for this ambiguity is required. This is done by numerically solving:

$$\varsigma = \min_{\mathbf{G}_j \subseteq \mathbf{SO_3}} \arccos\left[\mathrm{tr}\left(\frac{\mathbf{R}_i^\mathrm{T} \mathbf{G}_j - 1}{2}\right)\right] \tag{9}$$

where $\mathbf{G}_j$ is the subset of the group of rotation matrices $\mathbf{SO_3}$ that produces a diffraction equivalent orientation of $\mathbf{R}_j$. Given the principal value range of $\arccos(x)$ is confined between 0 and $\pi$, the misorientation between two diffractograms of a group is expected to follow a folded-normal distribution (Figure 10 right column). The misorientation distribution between recalibrated diffractograms is shown in Figure 10. The standard deviations derived from the folded normal distributions agree with the uncertainties derived from the single image orientation uncertainty (Figure 10D). This indicates that the diffraction incident beam direction has no significant orientation deviation over the imaging FOV. Notably, no significant improvement is observed constraining the energy to the initially calibrated value (Figure 10B vs. Figure 10C). Our results show the distribution of misorientations complies with a folded-gaussian distribution and it is bound within three times the standard deviation. This suggests the uncertainty in the detector orientation is in line with the uncertainty of the diffraction image in any position. Therefore, the beam direction can be assumed to be stable, and the moving beam configuration does not add any significant aberration to the accuracy of the diffraction instrument. This paves the way for a more common implementation of the moving beam diffraction geometry in new instruments.

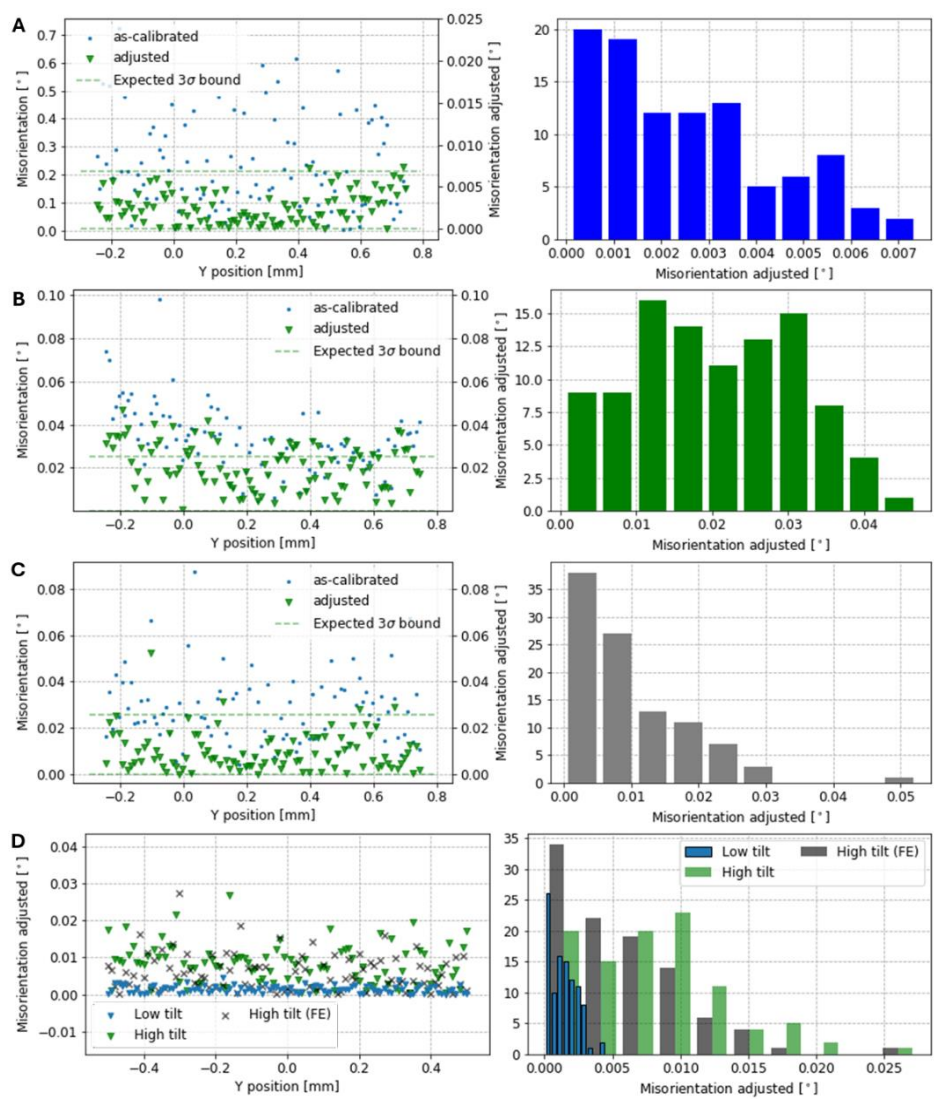

**Figure 10** **Detector misorientation.** Individual diffraction calibration misorientation (left column) and corresponding frequency distributions (right column). The measurements from a LaB$_6$ powder sample are repeated for different displacements of a capillary sample in the gravity direction while keeping the beam at constant KB-position and with the detector position (**A**) downstream normal-to- and (**B**, and **C**) inboard tilted-to-the-beam stream. The calibrations were repeated including the energy as a free degree-of-freedom (**A**, and **B**) or fixed (**C**, and **D**). In **D** are reported the misorientations measured with the same detector and sample configurations in **A** to **C**, but screening the y-axes with the KB-position motion while keeping the sample steady.

**3.2. Nearest-neighbour moving beam diffraction geometry calibration accuracy**

At DIAD, the data collected by the area detector is reduced to one-dimensional intensity profiles based on the diffraction geometry calibrated for the NIST standard sample in a flat-plate hollowed holder. A set of reference observations is collected with the diffraction beam stepping across the sample over a two-dimensional grid within the imaging FOV. The diffraction geometry of each observed diffractogram is independently calibrated. Data collected in routine experiments are then reduced utilizing the diffraction geometry of the nearest beam KB-position from the reference grid. Hence, the beam KB-position error propagates in the uncertainty of the geometry calibration.

We assess the nearest-neighbour calibration uncertainty by the error of the estimated interplanar distance for a diffraction ring crossing the middle of the detector panel relative to the average of those estimated with the reference self-calibrated diffraction geometry. We use the NIST SRM 660b $LaB_6$ standard interplanar distance for the 210 reflections, which is of ca. 1.85903 Å. Given the error is expected to increase with the distance between beam positions of the test and nearest-neighbour reference observation, test observations are collected on a grid shifted by half step compared to the reference one. Hence, we measure the uncertainty associated with the largest expected geometry error.

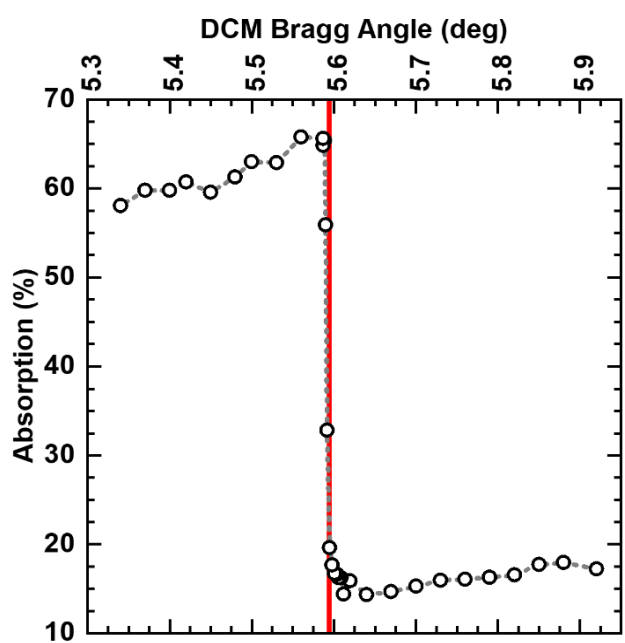

**Figure 11    Energy calibration of the diffraction beam via observation of a standard material absorption edge.** Intensity measurements were acquired using the imaging camera with a fixed Kirkpatrick–Baez (KB) mirror position. The large active area of the imaging scintillator enabled full beam capture, compensating for minor beam displacements caused by changes in the DCM configuration. Measurements were taken with and without the Mo foil in the beam path, allowing normalization of the transmitted intensity to account for flux variations across different Bragg angles. These variations arise from the energy-dependent efficiency of the DCM, KB mirrors, and the source emission profile.

We set the diffraction beam to a known energy to remove any source of uncertainty besides the diffraction geometry. We record the x-ray flux as a function of the DCM Bragg angle with the beam directly incident to the imaging camera and transmitted through a Mo foil. While the flux difference allows us to observe the Mo absorption energy (Figure 11), we tune the energy of the beam for the following experiment to ~20 keV by recovering the Bragg angle associated with the absorption edge. In addition to providing an independent energy calibration irrespective of the diffraction detector position, the close match between absorption-optimized Bragg angle and diffraction-estimated beam energy supports the reliability of the energy calibration performed with the detector in downstream normal-to-the-beam stream detector position. Notably, this experiment demonstrates also the possibility of collecting space-resolved X-ray absorption near-edge structure (XANS) information at DIAD.

Figure 12 shows the relative interplanar distance error estimated across the imaging FOV for different density reference calibration grids and the two detector positions downstream normal-to- and inboard tilted-to-the-beam stream. The error is directly dependent on the distance from the reference calibration beam positions. Indeed, the grid of the geometry calibration is clearly visible for the lower two density grids. As expected, based on the study of Figure 4, the downstream normal-to-the-beam stream detector position is characterized by larger errors compared to the inboard tilted-to-the-beam stream detector position. Notably, when the reference grid has a step size comparable to the beam spot size of 25 µm the interplanar distance does not show any evident correlation with the beam KB-position.

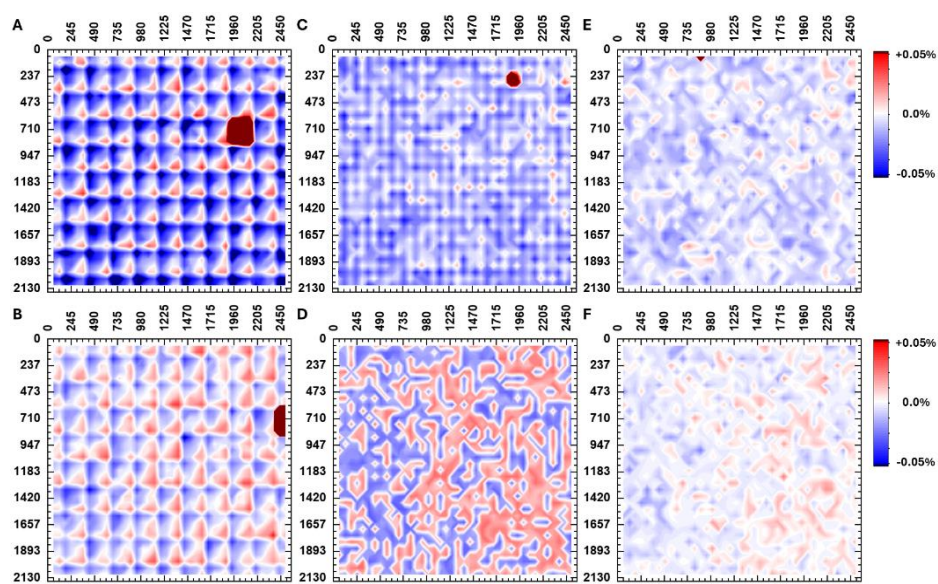

**Figure 12** **Nearest-neighbour moving beam geometry calibration error.** Interplanar distance error estimated with the nearest-neighbour calibration relative to the average of those estimated from the reference self-calibrated diffraction geometry for a flat-plate NIST SRM 660b $LaB_6$, mapped on the imaging field of view (2580×2180 pixels). The test diffraction data are reduced using the geometry calibrated for the nearest-neighbour standard diffraction from a regular cartesian array of 40×36 (**A**, and **B**), 20×18 (**C**, and **D**), and 10×9 (**E**, and **F**) beam offsets across the field-of-view. The diffraction beam moves across the imaging field-of-view on a cartesian grid with a half-step offset compared to the denser calibration array. The study is repeated with the detector positioned (**A**, **C**, and **E**) downstream normal-to-the-beam stream: i.e., $\alpha = 0°$, $\beta = 0°$, $\gamma = 0°$, and $r = 330$ mm; and at a standard experiment configuration (**B**, **D**, and **F**) downstream normal-to-the-beam stream shifted inboard to allow room for the imaging camera to be in view of the sample: i.e., detector panel orthogonal to the incident beam direction and the pixel at the center of the detector panel at $\alpha = \sim17°$, $\beta = \sim13°$, $\gamma = \sim4°$, and $r = 660$ mm. It is worth noting, the independent self-calibrated diffraction geometry yields a standard deviation of 0.008(1)% and ~0.006(5)% for the two detector configurations (Table 5), respectively.

**Table 5** Nearest-neighbour geometry calibration uncertainty.

|  | Detector position downstream normal-to-the-beam stream $\alpha = 0°, \beta = 0°, \gamma = 0°$, and $r = 330 mm$ | | Detector position inboard tilted-to-the-beam stream $\alpha = 17°, \beta = 13°, \gamma = 4°$, and $r = 660 mm$ | |
|---|---|---|---|---|
| average distance (Å) | 1.85903197 | | 1.85895051 | |
|  | relative error (%) | | relative error (%) | |
|  | average[*] | std | average[*] | std |
| self-calibrated | 0.006 | 0.008 | 0.005 | 0.006 |
| 40×36 | 0.008 | 0.010 | 0.005 | 0.007 |
| 20×18 | 0.015 | 0.015 | 0.020 | 0.021 |
| 10×9 | 0.027 | 0.032 | 0.015 | 0.038 |

[*] The averages are computed over the absolute relative errors.

The absolute error values remain below 0.01%, which is a limit minimum $d$-space change of $\sim 2e^{-4}$ Å. This is two orders of magnitude higher than the required strain accuracy from Withers 2004 (Withers, 2004b, 2004a). Indeed, it is the same accuracy we observed for self-calibrated datasets. Hence, a calibration grid of 40x36 diffraction spots evenly distributed across the imaging FOV ensures no aberration to the data reduction is introduced by the nearest-neighbour geometry calibration method. To compare this error with strain gauge resolution we need to consider their size. With a limit strain gauge resolution of 1μm/m and a typical size of 10 mm, the actual limit minimum displacement resolution is ~100 Å. Then, the same strain resolution is achieved for a strain gauge over a 1 mm length sample area, which is about the imaging FOV, and diffraction from a powder sample of 2 nm average size crystals, which is the minimum believable to produce Bragg peaks. This allows DIAD to measure mechanical strain information with comparable resolution at local and macroscopic scales.

## 4. Conclusions

Here we discuss the moving beam diffraction geometry implemented at the DIAD beamline at Diamond Light Source. Although scattering data is collected in transmission mode, the diffraction geometry is adjusted during experiments by changing the space location and possibly orientation of the incident beam. A rigid slit aperture and focusing mirror mounted on a motion stage allows the center of scattering at the sample across the imaging field of view. This enables the observation of space and time correlated structure information avoiding perturbation of the sample, which is kept stable. However, the dynamic change of the incident beam configuration requires the calibration of the relative detector position and orientation as function of any possible center of scattering. We achieve this *via* independent calibration of a dense two-dimensional grid of diffraction data sets collected from a flat-plate NIST standard calibrant, e.g., $LaB_6$ or $CeO_2$. Our measures confirm that diffraction is most sensitive to the moving geometry for the detector position downstream normal to the incident beam. Notably, the error induced by the moving beam is comparable to the effect of shifting the sample by an equivalent distance. Hence, the alternative option of moving the sample to screen the diffraction probe across it has no accuracy advantage compared to the moving beam geometry, whereas it will be affected by slower motions and possible perturbations of the sample macrostructure (e.g., when including a liquid bath environment). The observation data confirm the motion of the KB mirror coupled with a fixed aperture slit results in a rigid translation of the beam probe, without affecting the angle of the incident beam path to the sample. Here, we discuss the accuracy and precision of the beam stability and calibration process. We assess the reliability of the nearest-neighbour calibration method, where the data collected at any incident beam position is reduced assuming the detector position and orientation of the closest self-calibrated standard diffraction data. Our measures demonstrate that a nearest-neighbour calibration can achieve the same accuracy of a self-calibrated geometry when the distance between calibrated and probed sample region is smaller or equal to the beam spot size. We conclude that the nearest-neighbour calibration method yields an error comparable to the independent self-calibration. Notably, the error remains below 0.01%, which is a limit minimum d-space change of $\sim 2e^{-4}$ Å for a peak at Q of about 2 $Å^{-1}$. This proves the beamline achieves a suitable accuracy for measurement of structure distortion and peak shift of samples subject to change of phase and environmental conditions such as temperature and stress. Future work will further investigate the diffraction resolution for study of microstrain and phase decomposition, providing a characterization of the instrument broadening as function of the diffraction detector position and beam energy.

**Acknowledgements**   The authors are grateful to the Diamond Light Source (Experiment Number MG38775-1) for the provision of beamtime and for access to laboratory space.

**Appendix A. Diffraction detector pixel coordinates**

The detector plane is parametrized based on an origin point situated at the top-left corner of the top left-most pixel in the diffractogram. In the detector frame, the top-left corner position of the pixel in the $i^{th}$ column and $j^{th}$ row is:

$$\mathbf{p}_d = \begin{bmatrix} w & 0 \\ 0 & h \\ 0 & 0 \end{bmatrix} \cdot \begin{bmatrix} i \\ j \end{bmatrix}, \tag{A.1}$$

where, the pixel width, w, and height, h, are both 0.172 μm for the Pilatus2M CdTe in use at DIAD.

From the detector origin, $\mathbf{p}_0$, all other pixels, in the detector array are described according to the detector's orientation $\mathbf{R}_d$, by the transformation:

$$\mathbf{p} = \mathbf{p}_0 + \mathbf{R}_d \mathbf{p}_d, \tag{A.2}$$

The detector orientation is constructed from intrinsic composition of rotations about the three principal coordinate axes, beginning with yaw, defined negative anticlockwise around y, followed by a positive anticlockwise pitch rotation about x′, finally a roll negative anticlockwise around z″ is applied. Considering the canonical form of the right-handed rotation matrices about the principal axes by an angle $\theta$ to be $\mathbf{R}_x(\theta)$, $\mathbf{R}_y(\theta)$, and $\mathbf{R}_z(\theta)$, the composition is given by:

$$\mathbf{R}_d = \mathbf{R}_x(-\text{roll}) \cdot \mathbf{R}_x(\text{pitch}) \cdot \mathbf{R}_x(-\text{yaw}), \tag{A.3}$$

Taking the incident beam vector direction, $\hat{\mathbf{s}}_0$, as the $[0,0,1]^T$ of the laboratory system, the beam center is given by:

$$\mathbf{p}_{bc} = \mathbf{p}_0 + \mathbf{R}_d \mathbf{p}_{d,bc} = D \begin{bmatrix} 0 \\ 0 \\ 1 \end{bmatrix}, \tag{A.4}$$

**Appendix B. Scattering momentum error**

From Figure 3 $\mathbf{q} = \mathbf{s} - \mathbf{s}_0$ and $\mathbf{q}' = \mathbf{s}' - \mathbf{s}_0$. Hence, the scattering momentum error $\Delta \mathbf{q} = \mathbf{q}' - \mathbf{q}$ simplifies to

$$\Delta \mathbf{q} = \mathbf{s}' - \mathbf{s}. \tag{B.1}$$

Using Equations 5 and 6, Equation A.1 reads

$$\Delta \mathbf{q} = \frac{2\pi}{\lambda} \frac{\mathbf{p} - \boldsymbol{\delta}}{|\mathbf{p} - \boldsymbol{\delta}|} - \frac{2\pi}{\lambda} \frac{\mathbf{p}}{|\mathbf{p}|}. \tag{B.2}$$

This can be simplified with Equation 3 as,

$$\Delta \mathbf{q} = \frac{2\pi}{\lambda} \left\{ \frac{\hat{\mathbf{s}} \alpha_s - \boldsymbol{\delta}}{|\mathbf{p} - \boldsymbol{\delta}|} - \frac{\hat{\mathbf{s}} \alpha_s}{\alpha_s} \right\}, \tag{B.3}$$

$$\Delta \mathbf{q} = \frac{2\pi}{\lambda} \left\{ \left( \frac{\alpha_s}{|\mathbf{p} - \boldsymbol{\delta}|} - 1 \right) \hat{\mathbf{s}} - \frac{\boldsymbol{\delta}}{|\mathbf{p} - \boldsymbol{\delta}|} \right\}, \tag{B.4}$$

$$\Delta \mathbf{q} = \frac{2\pi}{\lambda} \frac{(\alpha_s - |\mathbf{p} - \boldsymbol{\delta}|)\hat{\mathbf{s}} - \boldsymbol{\delta}}{|\mathbf{p} - \boldsymbol{\delta}|}, \tag{B.5}$$

which is the same of Equation 7 in section 2.2.